\documentclass[aps,prd,amssymb,showpacs,twocolumn,superscriptaddress,nofootinbib,letterpaper]{revtex4}

\usepackage{setspace}
\usepackage{mathrsfs}

\usepackage{graphicx}
\usepackage{psfrag}
\usepackage{amsmath}
\usepackage{float}

\DeclareMathAlphabet{\mathpzc}{OT1}{pzc}{m}{it}

\def\dtm#1#2#3{\left( \begin{array}{ccc} {#1} &  0 & 0  \\ 0 & {#2} & 0 \\ 0 & 0 & {#3}  \end{array} \right) }

\def\ctm{\tilde{\gamma}_{ij}}

\def\ictm{\tilde{\gamma}^{ij}}
\def\ctc#1#2#3{\tilde{\Gamma}^{#1}_{#2 #3}}
\def\fctc#1#2#3{\mathring{\Gamma}^{#1}_{#2 #3}}

\def\DT#1{\tilde{D}_{#1}}
\def\UDT#1{\tilde{D}^{#1}}
\def\cec#1#2{\tilde{A}_{#1#2}}
\def\divb{\tilde{D}_k\beta^{k}}
\def\li{\mathscr{L}_{\vec{\beta}}}

\def\bi{\begin{itemize}}
\def\ei{\end{itemize}}
\def\be{\begin{equation}}
\def\ee{\end{equation}}
\def\bean{\begin{eqnarray}}
\def\eean{\end{eqnarray}}

\def\xf#1{\frac{\partial_r #1}{#1}}
\def\tf#1{\frac{\partial_t{#1}}{#1}}

\begin{document}
\title{Black hole critical behaviour with the generalized BSSN formulation}
\author{Arman Akbarian}
\affiliation{Department of Physics and Astronomy,
     University of British Columbia,
     Vancouver BC, V6T 1Z1 Canada}

\author{Matthew W. Choptuik}
\affiliation{CIFAR Cosmology and Gravity Program \\
     Department of Physics and Astronomy,
     University of British Columbia,
     Vancouver BC, V6T 1Z1 Canada}

\begin{abstract}
   The development of hyperbolic formulations of Einstein's equations has
   revolutionized our ability to perform long-time, stable, accurate numerical simulations
   of strong field gravitational phenomena.   However, hyperbolic methods have seen
   relatively little application in one area of interest, type II critical collapse, 
   where the challenges for a numerical code are particularly severe.  
   Using the critical collapse of a massless scalar field in spherical symmetry as a 
   test case, 
   we study a generalization of the Baumgarte-Shapiro-Shibata-Nakamura (BSSN) formulation
   due to Brown that is suited for use with curvilinear coordinates.  We adopt standard 
   dynamical gauge choices, including 1+log slicing and a shift that is either zero or 
   evolved by a Gamma-driver condition.  With both choices of shift we are able to evolve
   sufficiently close to the black hole threshold to (1) unambiguously identify the discrete 
   self-similarity of the critical solution, (2) determine an echoing exponent consistent 
   with previous calculations, and (3) measure a mass scaling exponent, also in accord with 
   prior computations.  Our results can be viewed as an encouraging first step towards
   the use of hyperbolic formulations in more generic type II scenarios, including the 
   as yet unresolved problem of critical collapse of axisymmetric gravitational waves.

\end{abstract}


\maketitle

\section{Introduction\label{sec:intro}}

In this paper we investigate the application of the
Baumgarte-Shapiro-Shibata-Nakamura (BSSN) formulation 
of Einstein's equations \cite{Shibata:1995,Baumgarte:1998}, as well as the
dynamical coordinate choices typically associated with it, within the context of 
critical gravitational collapse. 
The BSSN formulation
is a recasting of the standard 3+1 Arnowitt-Deser-Misner (ADM) \cite{ADM:1962} equations that is known to be strongly 
hyperbolic \cite{Sarbach:2002bt,Yoneda:2002kg} and suitable for numerical studies.
It has been widely used in numerical relativity and provides a robust and stable
evolution for the spacetime geometry. 
Most notably, various implementations
of this formulation have allowed successful computation of dynamical spacetimes
describing binaries of gravitationally-compact objects~\cite{Campanelli:2005dd, Baker:2005vv, CLP:2015}.
The standard gauge choices in BSSN---namely 
the 1+log slicing condition \cite{Bona:1997} and the Gamma-driver shift 
condition \cite{Alcubierre:2002}---are partial differential equations (PDEs) of  evolutionary type.
Furthermore, the BSSN approach results in a set of so-called free evolution equations, 
meaning that the Hamiltonian and momentum constraints are only solved at the initial time.
Thus, once initial data has been determined, one only has to solve time-dependent PDEs in order to 
compute the geometric variables in the BSSN scheme.  In particular, during the evolution there is no need 
to solve any elliptic equations, which in general could arise either from the constraints or from coordinate conditions. 
This is advantageous since it can be quite challenging to implement efficient numerical elliptic 
solvers.

In addition to the BSSN approach, the numerical relativity community has adopted 
the generalized harmonic (GH) \cite{Pretorius:2004} formulation of Einstein's equations,
which is also strongly hyperbolic and 
has performed very well in simulations of compact binaries \cite{Pretorius:2005gq, CLP:2015}.
Like BSSN, the GH formulation is of evolutionary type 
so that all of the metric components satisfy time-dependent PDEs.
It too uses dynamical coordinate choices: in this case one needs to provide a
prescription for the evolution of the harmonic functions defined by $H^{\mu}\equiv\Box x^{\mu}$.

Despite the tremendous success of these hyperbolic formulations 
in evolving strongly gravitating spacetimes containing black holes and neutron stars, 
they have not seen widespread use in another area
of strong gravity physics typically studied via numerical relativity, namely critical phenomena 
in gravitational collapse.
First reported in~\cite{Choptuik:1992} and briefly reviewed below,
critical phenomena emerge at the threshold of black hole formation
and present significant challenges for thorough and accurate computational treatment.
The original observation of critical behaviour as well as many 
of the subsequent studies were restricted to spherical symmetry
(for a review, see \cite{Gundlach:1997wm,Gundlach:2007lrr}) and there is a clear need 
to extend the work to more generic cases.  In this respect the BSSN and GH formulations 
would appear to be attractive frameworks.
However, it is not yet clear if these hyperbolic formulations, in conjunction with the 
standard dynamical gauge choices that have been developed, 
will allow the critical regime to be probed without the development of coordinate 
pathologies. 
Particularly notable in this regard is an implementation of the GH formulation that was employed
by Sorkin and Choptuik 
\cite{Sorkin:2009} to study the critical collapse 
of a massless scalar field in spherical symmetry.
Despite extensive experimentation with a variety of coordinate 
conditions, the code that was developed was not able to calculate near-critical
spacetimes: coordinate singularities invariably formed once the critical regime 
was approached.  
A natural question that then arises is whether the BSSN formulation (including 
the standard dynamical gauge choices used with it)
is similarly problematic or if it provides an effective
framework to study critical phenomena. 

Here we begin the task of addressing this question by revisiting the model of 
spherically symmetric massless scalar collapse. We use  a generalization of the BSSN 
formulation due to Brown~\cite{Brown:2009} that is well suited for use with curvilinear coordinates.
The choice of a massless scalar field as the matter source has the great advantage 
that the nature of the
critical solution is very well
known \cite{Gundlach:1995,PhysRevD.55.695,PhysRevD.55.R440,Garfinkle:1998,
PhysRevD.58.064031,MartinGarcia:2003gz}, making it straightforward for us to determine 
if and when our approach has been successful.
We note that although the calculations described below {\em are} restricted to spherical symmetry
our ultimate goal is to develop an
evolutionary scheme---including gauge choices---that can be applied to a variety of critical phenomena studies 
in axial symmetry and ultimately generic cases.

We now briefly review the main concepts and features of black hole critical phenomena that are most 
pertinent to the work in this paper.  Full details and pointers to the extensive
literature on the subject may be found in review articles 
 \cite{Gundlach:1997wm,Gundlach:2007lrr}.

Critical phenomena in gravitational collapse
can be described as a phase transition, analogous to that in a
thermodynamical system.
Under certain assumptions, 
a matter source
coupled to the Einstein gravitational field will evolve to one of two distinct final phases.
On the one hand, weak initial data will eventually disperse to infinity leaving
flat spacetime as the end state. On the other hand, sufficiently strong data will develop 
significant self gravitation and then collapse, resulting in a final phase 
which contains a black hole. 
Quite generically, remarkable behaviour emerges at and near
the transition between these phases, and this behaviour is precisely what we mean
by the critical phenomena in the system under consideration.

It transpires that there are two broad classes of critical phenomena that can be distinguished by
the behaviour of the black hole mass at threshold.  The class of interest here, known
as type II, is characterized by infinitesimal mass at the transition. 
Further, the black hole mass, $M_{\mathrm BH}$, satisfies a scaling law:
\be
M_{\mathrm BH} \sim |p-p^\star|^{\gamma} \, ,
\label{eq:massscale}
\ee
where $p$ is an arbitrary parameter that controls the strength of the
matter source at the initial time, $p^\star$ is the parameter value at threshold and
the mass scaling exponent,
$\gamma$, is a constant that is independent of the choice of the initial data.
Type II behaviour is also characterized by the emergence of a unique solution
at threshold which is generically self-similar. In some cases, including 
the massless scalar field, the self-similarity is discrete. 
Specifically, in spherically symmetric critical collapse with discrete self-similarity (DSS), as $p \to p^\star$
we find
\begin{equation}
Z^\star(\rho+\Delta,\tau+\Delta) \sim Z^\star(\rho,\tau) \, ,
\label{eq:echoing}
\end{equation}
where $Z^\star$ represents some scale-invariant component (function) of the critical solution.
\def\ms{{\mathrm S}}
Here
$\rho \equiv \ln(r_\ms)$ and $\tau \equiv \ln(T_\ms-T_\ms^\star)$ are logarithmically rescaled values of 
the areal radius,  $r_\ms$, and polar time, $T_\ms$, respectively, and
$T_\ms^\star$ is the accumulation time at which the central singularity associated 
with the DSS solution forms. $T_\ms$ has been normalized so that it measures proper time at the origin.
As with $\gamma$, the echoing (rescaling) exponent, $\Delta$, is a universal constant for a specific matter
source; i.e.~it is independent of the form of the initial data.

Another feature of type II collapse, intimately related to the self-similarity of the critical solution,
is that the curvature can become arbitrarily
large: in the limit of infinite fine-tuning, $p\to p^\star$, a naked singularity forms. 
Furthermore, the echoing behaviour (\ref{eq:echoing}) results in the development of 
fine structure in the solution
around the center of the scaling symmetry. Observing this structure and measuring the echoing
exponent $\Delta$ associated with it requires a code that can reliably evolve solutions very close to the critical
spacetime and that provides sufficient numerical resolution 
in the vicinity of the accumulation point $(r_\ms,T_\ms) = (0,T_\ms^\star)$.

As mentioned above, most studies of critical phenomena have assumed spherical symmetry.
This is particularly so for the case of type II behaviour where the resolution demands dictated
by the self-similarity of the critical solutions makes multi-dimensional work extremely 
computationally intensive.  As far as we know, the only work in spherical symmetry to have
used a purely evolutionary approach based on the BSSN or GH forms of the Einstein 
equations is~\cite{Sorkin:2009} which, as we have noted, was not successful in isolating the critical 
solution.\footnote{However, see~\cite{Pretorius:2000yu} for an investigation of type II behaviour in the 
collapse of a scalar field in 2+1 AdS spacetime that employs an ad hoc free evolution scheme.}
In axisymmetry 
there have been two investigations of type 
II collapse of massless scalar fields~\cite{Choptuik:2003ac, Choptuik:2004ha},
and several of type II collapse of pure gravitational waves 
(vacuum)~\cite{Abrahams:1993,Alcubierre:1999,Garfinkle:2000,Rinne:2008,Sorkin:2010}.
Of these, only Alcubierre et al.'s~\cite{Alcubierre:1999} and Sorkin's~\cite{Sorkin:2010} calculations of vacuum 
collapse adopted hyperbolic formalisms, 
and only the scalar field calculations---which employed a modified ADM formulation and
partially constrained evolution---were able to completely resolve the critical behaviour, including 
the discrete self-similarity of the critical solutions.   
In the fully three-space dimension (3D) context there have also been a few studies of 
type II collapse to date.  Perhaps most notable is the recent work of 
Healy and Laguna \cite{Healy:2013} which used a massless scalar
field as a matter source and the BSSN formulation with standard dynamical gauge choices. The
authors were 
able to observe the mass scaling~(\ref{eq:massscale}) with a measured $\gamma \approx 0.37$ consistent
with calculations in spherical symmetry.
However, they were not able to conclusively see the discrete self-similarity of the critical 
solution; in particular they could not accurately measure the echoing exponent, $\Delta$.
This shortcoming was attributed to a lack of computational resources rather
than a breakdown of the underlying methodology, including the coordinate conditions that were adopted.
Finally, there have been two attempts to probe the black hole threshold for the collapse
of pure gravitational waves in 3D~\cite{Santamaria:2006,Hilditch:2013}. Both employed a BSSN approach with,
for the most part, standard dynamical gauge choices. In both cases problems with the gauge apparently 
precluded calculation near the critical point (although resolution limitations may also have been an issue)
and neither the mass scaling nor the echoing exponent could be be estimated in either study.

We can thus summarize the state of the art in the use of hyperbolic formulations for the study 
of type II critical collapse as follows: 
to our knowledge there has been no implementation of a fully 
evolutionary scheme, based on either BSSN or GH, that has allowed for evolution
sufficiently close to a precisely critical solution to allow the unambiguous identification of discrete 
self-similarity (or continuous self-symmetry for that matter).  Again, and particularly in light of 
the experience of~\cite{Sorkin:2009}, the key 
aim of this paper is to investigate the extent to which it {\em is} possible to use a 
BSSN scheme to fully resolve type II solutions.   A major concern here is the appropriate
choice of coordinate conditions, not least 
since 
dynamical
gauge choices can be prone to the development of gauge shocks and other 
types of coordinate singularities~\cite{Alcubierre:1996,Alcubierre:2002i}. 
Such pathologies 
could, in principle, prevent a numerical solver from evolving the spacetime in or near 
the critical regime.

Now, as Garfinkle and Gundlach have discussed in detail~\cite{Garfinkle:1999cm}, an ideal coordinate 
system for numerical studies of type II collapse is one which {\em adapts} itself to the 
self-similarity: for the DSS case this means that the metric coefficients and relevant matter 
variables are exactly periodic in the coordinates in the fashion given by~(\ref{eq:echoing}).
Clearly, if the coordinate system {\em is} adapted, then other than at the naked singularity---which
is inaccessible via finite-precision calculations---it should
remain non-singular during a numerical evolution. One can then argue that ensuring that the numerical 
scheme has adequate resolution will be the key to successful simulation of the critical 
behaviour.  At the same time, it is also clear that there will be coordinate systems which do not 
necessarily adapt
but which nonetheless remain non-singular during critical collapse, at least over some range
of scales, and which are therefore potentially useful for numerical calculations.  We will see below 
that there is strong evidence that the coordinate systems we have used belong to the latter class, 
and weaker evidence that they {\em do} adapt to the self-similarity.

Another potential source of problems, which is not specific to hyperbolic 
formulations, relates to our restriction to spherical symmetry.  As is 
well known, the singular points of curvilinear coordinate systems, $r=0$
in our case, can sometimes require special treatment to ensure that
numerical solutions remain regular there.  In critical collapse the highly dynamical
nature of the solution near $r=0$ might naturally be expected to exacerbate 
problems with regularity.
In the work described below we have paid special attention to the ability of our approach
to both fully resolve the near-critical configuration and maintain regularity of the solution
at the origin.

The remainder of this paper is organized as follows: in Sec.~\ref{sec:EOM} we 
review the generalized BSSN formulation and display the equations of motion for 
our model system. 
Sec.~\ref{sec:numerics} 
expands the discussion of the issue of regularity at the coordinate
singularity point, describes the numerical approach we have adopted, and provides details
concerning the 
various tests and diagnostics we have used to validate our implementation.
In Sec.~\ref{sec:results} we present 
results computed using two distinct choices for the shift vector
and provide conclusive
evidence that the generalized BSSN formulation is capable of evolving 
in the critical regime in both cases.
Sec.~\ref{sec:conclusion} 
contains some brief concluding remarks, and further details concerning the BSSN formalism
in spherical symmetry and the scalar field equations of motion are included 
in App.~\ref{sec:appdx1} and App.~\ref{sec:appdx2}, respectively.
We adopt units where 
the gravitational constant and the speed of light are both unity:
$G=c=1$.

\section{Equations of Motion\label{sec:EOM}}
The dynamical system we intend to study in the critical collapse regime is
a real, massless scalar field, $\Psi$, self gravitating via Einstein's equations,
\begin{equation}
G_{\mu\nu} = 8 \pi T_{\mu\nu} \, .
\label{eq:einsteinequatoin}
\end{equation}
Here, $T_{\mu\nu}$ is the energy-momentum tensor associated with
the minimally coupled $\Psi$:
\be
T_{\mu\nu} = \nabla_{\mu}\nabla_{\nu}\Psi -\frac{1}{2}g_{\mu\nu}\nabla^{\eta}\Psi\nabla_{\eta}\Psi \, ,
\ee
and the evolution of the scalar field is given by
\be
\nabla^\mu \nabla_\mu \Psi = 0 \, .
\label{eq:sce}
\ee
The time-development of the geometry is then given by recasting Einstein's equations
as an evolution system based on the usual 3+1 expression for the spacetime metric:
\begin{equation}
	ds^2 = -\alpha^2 dt^2 + \gamma_{ij}(dx^i+\beta^i dt)(dx^j+\beta^j dt) \, .
\end{equation}
Here, the 3-metric components, $\gamma_{ij}$, are viewed as the 
fundamental dynamical geometrical variables 
and the lapse function, $\alpha$, and shift vector,
$\beta^i$, 
which encode 
the coordinate freedom of general relativity, must in general 
be prescribed independently of the equations of motion.

\subsection{Generalized BSSN \label{sec:EOMgeometry}}

%
We now summarize the BSSN formulation of Einstein's equations and describe
how it can be adapted to curvilinear coordinates. Readers interested in additional details 
are directed to 
\cite{BS:2010} 
for a more pedagogical discussion.

In the standard ADM formulation \cite{ADM:1962, York:1978}, the dynamical Einstein equations
are rewritten as evolution equations for the 3-metric and the
extrinsic curvature $\{\gamma_{ij},K_{ij}\}$. 
The first difference between the BSSN formulation and 
the ADM decomposition is the conformal re-scaling of the ADM dynamical
variables:
\begin{equation}
	\gamma_{ij} = e^{4\phi}\ctm \, ,
	\label{eq:decomgam}
\end{equation}
\begin{equation}
	K_{ij} = e^{4\phi}\cec{i}{j} + \frac{1}{3}\gamma_{ij}K \, , 
\end{equation}
where $e^{\phi}$ is the conformal factor, $\ctm$ is the conformal metric,
$\cec{i}{j}$ is the conformally rescaled trace-free part of the 
extrinsic curvature and $K=\gamma^{ij}K_{ij}$ is the
trace of the extrinsic curvature. Here by 
fixing the trace of $\tilde{A}_{ij}$, and the determinant of the
conformal metric, the set of primary ADM dynamical variables
transforms to the new set:
\be
\{\gamma_{ij},K_{ij}\} \rightarrow \{\phi,\ctm,K,\cec{i}{j}\} \, ,
\ee
in the BSSN formulation.

In the original BSSN approach,
the conformal metric $\ctm$ is taken
to have determinant $\tilde{\gamma} = 1$.
However this choice
is only suitable when we adopt coordinates in which the 
determinant of the flat-space metric 
reduces to unity. 
This is the case, of course, for Cartesian coordinates but is not so for general
curvilinear systems.
For instance, the
flat 3-metric in spherical coordinates:
\be
ds^2 = dr^2 + r^2 d\theta^2 + r^2\sin^2\theta d\phi^2 \, ,
\ee
has determinant 
$\mathring{\gamma}=r^4\sin^2\theta$. 
Recently, Brown \cite{Brown:2009} has resolved this issue 
by introducing a covariant version of the BSSN equations---the so-called {\em generalized} BSSN formulation,
which we will hereafter refer to as G-BSSN---in which the primary dynamical
variables are tensors so that the formulation can be adapted to non-Cartesian 
coordinate systems.
In G-BSSN we no longer assume that
the conformal 3-metric has determinant one. 
Rather, $\phi$ becomes a true scalar 
and for its dynamics to be determined a prescription for the time 
evolution of the determinant of $\ctm$ must be given. 
In the following this will be done by requiring that the determinant be 
constant in time.

Another main difference between the ADM decomposition and BSSN is that 
the mixed spatial derivative terms occurring in the 3-Ricci
tensor are eliminated through the definition of a new quantity, $\tilde{\Gamma}^{k}$:
\begin{equation}
\tilde{\Gamma}^{k} \equiv \ictm \ctc{k}{i}{j} \, ,
\end{equation}
which becomes an additional, independent dynamical variable. Note that 
$\tilde{\Gamma}^i$ is not a vector as it is coordinate dependent.
To extend this redefinition so that it is well suited for all coordinate choices,
in G-BSSN we define
\begin{equation}
	\tilde{\Lambda}^{k} \equiv \ictm (\ctc{k}{i}{j} - \fctc{k}{i}{j}) = \tilde{\Gamma}^k - \fctc{k}{i}{j}\ictm \, ,
	\label{eq:deflam}
\end{equation}
where $\fctc{k}{i}{j}$ denotes the Christoffel symbols associated with the flat metric. 
This definition makes this so-called conformal connection, $\tilde{\Lambda}^i$, a true vector 
and it becomes a primary dynamical
variable in G-BSSN. 

We now summarize the G-BSSN equations, referring
the reader to
\cite{Alcubierre:2010} for more details, including a full derivation.
We begin by defining $\partial_{\bot}$, the time derivative operator acting normally
to the $t={\rm const.}$ slices:
\begin{equation}
	\partial_{\bot} \equiv \partial_{t} - \li \, ,
\end{equation}
where $\li$ denotes the Lie derivative along ${\vec \beta}$.
We then have
\begin{equation}
	\partial_{\bot} \phi = -\frac{1}{6} \alpha K + \sigma\frac{1}{6}\divb \, ,
	\label{eq:evolphi}
\end{equation}
\begin{equation}
	\partial_{\bot} \ctm = -2\alpha\cec{i}{j} - \sigma \frac{2}{3} \cec{i}{j} \divb \, ,
\label{eq:evolctm}
\end{equation}
\begin{equation}
	\partial_{\bot} K = -\gamma^{ij}D_j D_i\alpha + \alpha(\tilde{A}_{ij}\tilde{A}^{ij}+\frac{1}{3}K^2) + 4\pi(\rho+S) \, ,
\end{equation}
\begin{eqnarray}
	\partial_{\bot}\tilde{A}_{ij} &=& e^{-4\phi}\left[ -D_i D_j \alpha + \alpha(R_{ij} - 8\pi S_{ij}) \right]^{\mathrm{TF}} \nonumber \\
 &+& \alpha(K\tilde{A}_{ij} - 2\tilde{A}_{il}\tilde{A}^l_{\;j}) - \sigma\frac{2}{3}\tilde{A}_{ij}\divb \, ,
	\label{eq:evolA}
\end{eqnarray}
\begin{eqnarray}
	\partial_{\bot} \tilde{\Gamma}^{i} &=& -2\tilde{A}^{ij} \partial_{j} \alpha + \tilde{\gamma}^{lj} \partial_{j}\partial_{l}\beta^{i} \nonumber \\
&+& 2\alpha \left( \ctc{i}{j}{k}\tilde{A}^{kj} -\frac{2}{3} \ictm \partial_{j} K + 6 \tilde{A}^{ij} \partial_{j} \phi -8\pi \tilde{\gamma}^{ij}S_j \right) \nonumber \\
&+& \frac{\sigma}{3}\left[ 2 \tilde{\Gamma} ^{i} \divb + \tilde{\gamma}^{li} \partial_{l}(\divb) \right] \, .
	\label{eq:evolgamma}
\end{eqnarray}
Here, a superscript TF denotes the trace-free part
(with respect to the 3-metric $\gamma_{ij}$) of a tensor, and 
$\tilde{D}_i$ is the covariant derivative associated with the conformal metric $\ctm$.
Additionally, the quantity $\sigma$ is an adjustable parameter that is discussed below
and typically
is either 0 or 1.
Note that all the Lie derivatives in the G-BSSN equations 
operate on true tensors and vectors of weight 0. For instance,
\begin{eqnarray}
\mathscr{L}_{\vec{\beta}} \tilde{A}_{ij} = \beta^{k} \partial_{k} \tilde{A}_{ij} + \cec{i}{k} \partial_{j} \beta^{k} + \cec{k}{j} \partial_{i} \beta^{k} \, .
\end{eqnarray}
Furthermore, in G-BSSN, rather than evolving (\ref{eq:evolgamma}), the redefined 
conformal connection, $\tilde{\Lambda}^i$, is evolved via
\begin{equation}
	\partial_{t} \tilde{\Lambda}^{k} = \partial_{t} \tilde{\Gamma}^{k} - \fctc{k}{i}{j} \partial_{t} \ictm \, ,
	\label{eq:evollam}
\end{equation}
where the time derivative $\partial_{t} \ictm$ is eliminated using
(\ref{eq:evolctm}).
In equation (\ref{eq:evolA}), $R_{ij}$ denotes the 3-Ricci tensor
associated with $\gamma_{ij}$ and can be written as the sum
\begin{equation}
	R_{ij} = R^{\phi}_{ij} + \tilde{R}_{ij} \, ,
\end{equation}
where $R^{\phi}_{ij}$ is given by
\begin{equation}
R^{\phi}_{ij} = -2\DT{i} \DT{j} \phi - 2 \ctm \UDT{k} \DT{k} \phi + 4 \DT{i} \phi \DT{j} \phi - 4 \ctm \UDT{k} \phi \DT{k} \phi \, ,
\end{equation}
and $\tilde{R}_{ij}$ is the 3-Ricci tensor associated with the conformal metric:
\begin{eqnarray}
	\tilde{R}_{ij} = &-&\frac{1}{2} \tilde{\gamma}^{lm}\partial_{m}\partial_{l}\ctm + \tilde{\gamma}_{k(i}\partial_{j)}\tilde{\Gamma}^{k} + \tilde{\Gamma}^{k}\tilde{\Gamma}_{(ij)k} \nonumber \\
&+& \tilde{\gamma}^{lm}\left( 2\ctc{k}{l}{(i} \tilde{\Gamma}_{j)km}  + \ctc{k}{i}{m} \tilde{\Gamma}_{klj}  \right) \, .
\end{eqnarray}
The matter fields $\rho$, $S$, $S_i$ and $S_{ij}$ are defined by

\be
\rho = n_\mu n_\nu T^{\mu\nu} \, ,
\label{eq:defrho}
\ee
\be
S = \gamma^{ij}S_{ij} \, ,
\ee
\be
S^i = -\gamma^{ij}n^\mu T_{\mu j} \, ,
\ee
\be
S_{ij} = \gamma_{i\mu}\gamma_{i\nu}T^{\mu\nu} \, ,
\ee
where $n^\mu$ is the unit normal vector to the $t=\mathrm{const.}$ slices.

As mentioned previously, we need to prescribe dynamics for the determinant of $\ctm$ to
have a complete set of equations of motion for the  G-BSSN dynamical variables.
One approach is to fix the determinant to its
initial value by demanding that
\be
\partial_t \tilde{\gamma}=0 \, .
\label{eq:lagchoice}
\ee
This is the so-called Lagrangian option
and is associated with the choice $\sigma=1$ in the equations. Another option is to
define the determinant to be constant along the normal direction to the 
time slices, which can be implemented by requiring $\partial_{\bot} \tilde{\gamma} =0$. This is usually referred to as the
Lorentzian option, and is associated with the choice $\sigma=0$.
Here we choose (\ref{eq:lagchoice}), i.~e.~$\sigma=1$.

Note that in the G-BSSN equations the divergence of the shift vector,
\begin{equation} 
\divb = \frac{1}{\sqrt{\tilde{\gamma}}}\partial_{k}(\sqrt{\tilde{\gamma}}\beta^k) \, ,
\end{equation}
no longer reduces to $\partial_{k}\beta^k$ since the 
determinant of the conformal metric $\tilde{\gamma}_{ij}$ is not necessarily 1, but by virtue of
the choice (\ref{eq:lagchoice}) is
equal to that of the initial background flat metric in the chosen curvilinear coordinates. 

As usual, when setting initial data for any given evolution of the coupled Einstein-matter equations
we must solve the Hamiltonian and momentum constraints.  
In terms of the G-BSSN variables these are
\begin{eqnarray}
   \mathscr{H} &\equiv& \tilde{\gamma}^{ij}\DT{i}\DT{j}e^\phi - \frac{e^\phi}{8}\tilde{R} + \frac{e^{5\phi}}{8}\tilde{A}^{ij}\tilde{A}_{ij} \nonumber \\
	&-& \frac{e^{5\phi}}{12}K^2 + 2\pi e^{5\phi}\rho  = 0\, ,
	\label{eq:HC}
\end{eqnarray}
\begin{equation}
	\mathscr{M}^i \equiv \tilde{D}_j\left(e^{6\phi}\tilde{A}^{ji}\right) - \frac{2}{3}e^{6\phi}\tilde{D}^i K - 8\pi e^{6\phi} S^i = 0 \, .
\label{eq:MC}
\end{equation}



\subsection{G-BSSN in spherical symmetry and gauge choices \label{sec:spherical}}
In spherical symmetry a generic form of the conformal metric $\ctm$ is given by

\def\A{\tilde{\gamma}_{rr}}
\def\B{\tilde{\gamma}_{\theta\theta}}
\def\tA{\tilde{A}}
\def\tLambda{\tilde{\Lambda}}
\begin{equation}
	\ctm = \dtm{\A(t,r)}{r^2\B(t,r)}{r^2 \B(t,r)\sin^2\theta} \, .
	\label{eq:gammaspher}
\end{equation}
Similarly, a suitable ansatz for the traceless extrinsic curvature is
\begin{equation}
	\tA_{ij} = \dtm{\tA_{rr}(t,r)}{r^2\tA_{\theta\theta}(t,r)}{r^2\tA_{\theta\theta}(t,r)\sin^2\theta } \, .
   \label{eq:Aspher}
\end{equation}
The shift vector and $\tilde{\Lambda}^i$ have only radial components:
\begin{equation}
	\beta^i = \left[\beta(t,r),0,0\right] \, ,
\end{equation}
\begin{equation}
	\tilde{\Lambda}^i = \left[\tilde{\Lambda}(t,r),0,0\right] \, .
   \label{eq:vecspher}
\end{equation}
Given~(\ref{eq:gammaspher}-\ref{eq:vecspher}),
the G-BSSN equations become a set of first order evolution equations for the 7
primary variables
\begin{eqnarray}
\Big\{\phi(t,r),\A(t,r),\B(t,r),K(t,r), \nonumber \\
\tA_{rr}(t,r),\tA_{\theta\theta}(t,r),\tLambda(t,r) \Big\} \nonumber \, .
\end{eqnarray}
These are coupled to the evolution equation~(\ref{eq:sce}) 
for the scalar field 
and 
constrained by the initial conditions (\ref{eq:HC}--\ref{eq:MC}).
The explicit expressions for the full set of equations of motion are given in 
App.~\ref{sec:appdx1}.

To fix the time slicing we implement a non-advective\footnote{
The terminology {\em non-advective} derives from the
absence of an ``advective'' term, $\beta^j\partial_j$, on the 
left hand side of equations (\ref{eq:1pl},\ref{eq:gammadrive}). we note
that we also experimented with the advective versions of the equations. the results were 
very similar to those for the non-advective case; in particular, near-critical solutions
exhibiting echoing and scaling could also be obtained.
}
version of the 1+log slicing condition:\footnote{
The reader can easily check that in the case of zero shift, the lapse 
choice given by~(\ref{eq:1pl})
combined with (\ref{eq:evolphi}) implies
$\partial_t ( \alpha - \phi/12 ) = 0$. in Cartesian coordinates $\phi/12 \equiv \ln\gamma$, so
this last equation gives $\alpha - \ln\gamma = c(\vec{x})$, where the
function $c(\vec{x})$ is time independent. the choice $c(\vec{x})=1$ then yields an algebraic expression for the
lapse, $\alpha = 1 + \ln\gamma$, which  is the origin of the 
terminology ``1+log slicing''.
}
\begin{equation}
	\partial_{t} \alpha = -2\alpha k \, .
	\label{eq:1pl}
\end{equation}
for the spatial coordinates we either choose a zero shift:
\begin{equation}
   \beta^i = 0 \, ,
\end{equation}
or  use what we will term the gamma-driver condition:
\begin{equation}
	\partial_t \beta^i = \mu \tilde{\Lambda}^i -\eta\beta^i \, .
	\label{eq:gammadrive}
\end{equation}
Here, $\mu$ and $\eta$ are adjustable parameters which we set to $\mu=3/4$ and 
$\eta\simeq1/(2M_{\rm ADM})$, where $M_{\rm ADM}$ is the total mass
of the system measured at infinity (see~Sec.~\ref{sec:constraints_conserved}).
We emphasize that~(\ref{eq:gammadrive}) is {\em not} the usual Gamma-driver equation used in the 
standard BSSN approach:
\begin{equation}
	\partial_t \beta^i = \mu \tilde{\Gamma}^i -\eta\beta^i \, ,
	\label{eq:gammadrive-real}
\end{equation}
but since it is a natural extension of the above to the G-BSSN case we have opted
to use the same nomenclature.
\newcommand{\betagamma}{\beta^{G}}
In the rest of this paper, we frequently refer to the shift 
vector evolved via~(\ref{eq:gammadrive}) as $\betagamma$.
Explicitly, in spherical symmetry $\betagamma$ is defined by
\be
\partial_t \betagamma(t,r) =  \mu \tilde{\Lambda}(t,r) -\eta\betagamma(t,r) \, .
\label{eq:betagammadrive}
\ee

\section{Numerics \label{sec:numerics}}

We use a second order finite differencing method to discretize equations 
(\ref{eq:evolphi}-\ref{eq:evolA}) and (\ref{eq:evollam}). Further, the
equations of motion are transformed to a compactified radial coordinate that 
we denote by ${\tilde r}$ and which is defined in terms of the original
coordinate $r$ by
\begin{equation}
	r = e^{\tilde{r}} - e^{\delta} + \frac{R_{\infty}}{R_{\infty}-\tilde{r}} - \frac{R_{\infty}}{R_{\infty}-\delta} \, ,
	\label{eq:compactify}
\end{equation}
where $\delta$ and $R_\infty$ are parameters with 
typical values $\delta\simeq-12$ and $R_\infty\simeq3$. 
It is straightforward to verify the following: (1)
the radial domain $r=(0,\infty)$ maps 
to the computational domain ${\tilde r}=(\delta,R_\infty)$,
(2) the derivative ${dr}/{d\tilde{r}}$ decreases
toward the origin ($\tilde{r} \simeq \delta$), so that a uniform grid on $\tilde{r}$ 
is a non-uniform grid on $r$ with approximately $10^3$ times more 
resolution close to the origin relative to the outer portion of the solution
domain,
$\tilde{r}\simeq 2$ ($r\simeq10$), where the support of the scalar field is 
initially concentrated,
(3) the parameter $\delta$ can be used to adjust the resolution near
the origin; specifically, decreasing $\delta$ increases 
the resolution near $r=0$. 
For notational simplicity, however, in the following we omit the explicit 
dependence of the fields on $\tilde{r}$ and denote the spacetime dependence of
any dynamical variable $X$ as previously: $X\left(t,r(\tilde{r})\right) \equiv X(t,r)$. 

We use a finite difference grid that is  uniform in $\tilde{r}$ and
analytically transform all $r$-derivative terms in the equations of motion 
to their $\tilde{r}$-coordinate counterparts prior to finite-differencing.

We also developed a Maple-based toolkit \cite{akbarian:2014fd} 
that automates the process of discretizing
an arbitrary derivative expression. 
This toolkit handles boundary conditions
and generates a point-wise Newton-Gauss-Seidel 
solver in the form of Fortran routines for a given set of time 
dependent or elliptic PDEs . The calculations in this paper were all 
carried out using this infrastructure. 

\subsection{Initialization\label{sec:initialization}}
The matter content is set by initializing the scalar field to a localized Gaussian shell:
\be
\Psi(0,r) = p\exp\left( -\frac{(r-r_0)^2}{\sigma_r^2} \right) \, ,
\label{eq:initpsi}
\ee
where $p$, $r_0$ and $\sigma_r$ are parameters.
Note that here $r$ is the non-compactified radial coordinate which is 
related to the compactified coordinate $\tilde{r}$ via~(\ref{eq:compactify}).
A typical initial profile
for the scalar field in our calculations has $\sigma_r\simeq1$, 
$r_0\simeq 10$, and $p$ of order $10^{-1}$. We use the overall amplitude
factor $p$ as the tuning parameter to
find critical solutions.
We initialize the conformal metric (\ref{eq:gammaspher}) to the flat metric in spherical
symmetry,
\be
\A(0,r)=\B(0,r) = 1 \, ,
\ee
and initialize the lapse function to unity,
\be
\alpha(0,r) = 1 \, .
\ee
We also demand that the initial data be time-symmetric,
\be
\tA_{rr}(0,r) = \tA_{\theta\theta}(0,r) = K(0,r) =  0 \, ,
\ee
\be
\beta(0,r) = \tLambda(0,r) = 0 \, ,
\ee
\be
\partial_t \Psi(t,r)|_{t=0} = 0 \, ,
\ee
which means that the momentum constraint~(\ref{eq:MC}) is trivially satisfied.  This leaves
the Hamiltonian constraint~(\ref{eq:HC}) 
which is solved as a two-point boundary value problem for the conformal factor
at the initial time,
\be
\psi(r) \equiv e^{\phi(0,r)} \, .
\ee
The outer boundary condition for $\psi$,
\be
\psi(r) |_{r=\infty} = 1 \, ,
\label{eq:psiinf}
\ee
follows from asymptotic flatness, while at $r=0$ we have
\be
\partial_r \psi(r) |_{r=0} = 0 
\ee
since $\psi(r)$ must be an even function in $r$ for regularity at the 
origin.
\subsection{Boundary Conditions}
Due to the fact that the metric has to be conformally flat at the origin we have
\be
\A(t,0)=\B(t,0) \, .
\ee
Further, since we are using the Lagrangian choice, $\sigma=1$, 
the determinant of $\ctm$ must at all times be equal to its value at the initial time, 
so 
\be
\A\B^2=1 \, .
\label{eq:detgamone}
\ee
From these two results we have
\be
\A(t,0) = \B(t,0) = 1 \, .
\label{eq:bdab}
\ee
Using~(\ref{eq:bdab}) and (\ref{eq:evolctm}) it is then easy to see that we must also have
\be
\tA_{rr}(t,0) = \tA_{\theta\theta}(t,0) = 0 \, .
\label{eq:bdAA}
\ee

As is usual when working in spherical coordinates, many of the boundary conditions that 
must be applied at $r=0$ follow from the demand that the solution be regular there. 
Essentially, the various dynamical variables must have either even or odd ``parity''
with respect to expansion in $r$ as $r\to0$.  Variables with even parity, typically
scalars 
%
or diagonal components of rank-2 tensors, must have vanishing radial derivative at
$r=0$, while odd parity functions, typically radial components of vectors, will
themselves vanish at the origin.

Applying these considerations to our set of unknowns we find
\be
\partial_r \A(t,r)|_{r=0} = \partial_r \B(t,r)|_{r=0} = 0 \, ,
\label{eq:bdabp}
\ee
\be
\partial_r \tA_{rr}(t,r)|_{r=0} = \partial_r \tA_{\theta\theta}(t,r)|_{r=0} = 0 \, ,
\label{eq:bdAAp}
\ee
\be
\beta(t,0) = \tLambda(t,0) = 0 \, ,
\label{eq:bdbeta}
\ee
\be
\partial_r K(t,r)|_{r=0} = \partial_r \phi(t,r)|_{r=0} =  \partial_r \Psi(t,r)|_{r=0}= 0 \, .
\label{eq:bdK}
\ee
We use equations 
(\ref{eq:bdab},\ref{eq:bdAA},\ref{eq:bdbeta})
to fix the values of the functions at the origin and a forward 
finite-differencing of (\ref{eq:bdK}) to update $K$, $\phi$
and $\Psi$
at $r=0$. 
Further, we apply a forward finite-differencing of 
(\ref{eq:bdabp},\ref{eq:bdAAp}) to update the 
values of the function at the grid point next to the origin.
The 1+log condition (\ref{eq:1pl}) can be used directly at $r=0$.
Again, we emphasize that all of the $r$-derivative terms of the
boundary conditions described above
are analytically transformed to the numerical coordinate, $\tilde{r}$, before
the equations are finite-differenced.

Since we are using compactified coordinates,
all the variables are set to their flat spacetime values at the outer boundary
$r=\infty$:
\be
\A = \B = e^\phi = \alpha = 1 \;\;\; \mathrm{at:} \; (t,\infty) \, ,
\ee
\be
\tA_{rr} = \tA_{\theta\theta} = K = \tLambda = \beta = \Psi = 0 \;\;\; \mathrm{at:} \; (t,\infty) \, .
\ee
Here, we emphasize that spatial infinity, $r=\infty$, corresponds to the 
finite compactified (computational) coordinate point $\tilde{r} = R_\infty$.

\subsection{Evolution scheme and regularity\label{sec:evolution}}

We implemented a fully implicit, 
Crank-Nicolson \cite{CN:1996} finite differencing
scheme to evolve the system of G-BSSN equations. 
The precise form of the continuum equations used is given in App.~\ref{sec:appdx1} and 
all derivatives, both temporal and spatial,
were approximated using second-order-accurate finite difference
expressions.  

During an evolution
the correct limiting behaviour 
of the spatial metric components 
must be maintained near $r=0$ 
to ensure a regular solution. 
For example, the limiting values
of the conformal metric components $\A$ and $\B$ are given by

\be
\A(t,r) = 1 + O(r^2) \, ,
\label{eq:alimitr}
\ee
\be
\B(t,r) = 1 + O(r^2) \, .
\label{eq:blimitr}
\ee
If the discrete approximations
of the metric functions do not satisfy these conditions, then irregularity
will manifest itself in the divergence of 
various expressions such as the Ricci tensor component (\ref{eq:riccitenrr})
\be
R_{rr} = 2\frac{\A-\B}{r^2\B} + \cdots \, ,
\ee
which should converge to a finite value at the origin
if conditions (\ref{eq:alimitr},\ref{eq:blimitr}) hold.

One approach to resolve potential regularity issues
is to regularize the equations 
\cite{Ruiz:2007,Alcubierre:2010,Sorkin:2009},
by redefining the
primary evolution variables, so that 
the equations become manifestly regular at the origin. Another approach is to use
implicit or partially implicit methods \cite{CorderoCarrion:2012}. As recently shown
by Montero and Cordeo-Carrion \cite{Montero:2012}, such schemes 
can yield stable evolution without need for explicit regularization.
Baumgarte et~al.~\cite{Baumgarte:2012} also adopted a similar approach---using a partially
implicit scheme without regularization---in an implementation of 
the G-BSSN formulation in spherical polar coordinates.

As mentioned, our implementation is fully implicit and
we have also found that our
generalized BSSN equations can be evolved without any need for regularization
at the origin, even in strong gravity scenarios where the spacetime 
metric has significant deviations from flatness near the origin.

That said, we also experimented with other techniques aimed at improving regularity.
For example, using the constraint equation
(\ref{eq:detgamone}) and the fact that $\tA_{ij}$ is trace-free,
\be
\frac{\tilde{A}_{rr}}{\A}+2\frac{\tilde{A}_{\theta\theta}}{\B} = 0 \, ,
\label{eq:tracea}
\ee
we can compute $\B$  and
$\tA_{\theta\theta}$  
in terms of 
$\A$ and $\tA_{rr}$, 
respectively,
rather than evolving them. 
However, when we did this we found 
no significant improvement in regularity relative to the original scheme.

Finally, to ensure our solutions remain smooth on the scale of the mesh 
we use fourth order Kreiss-Oliger dissipation~\cite{KO:1973} 
in the numerical solution updates.

\begin{figure}
	\begin{center}
		\includegraphics[width=0.42\textwidth]{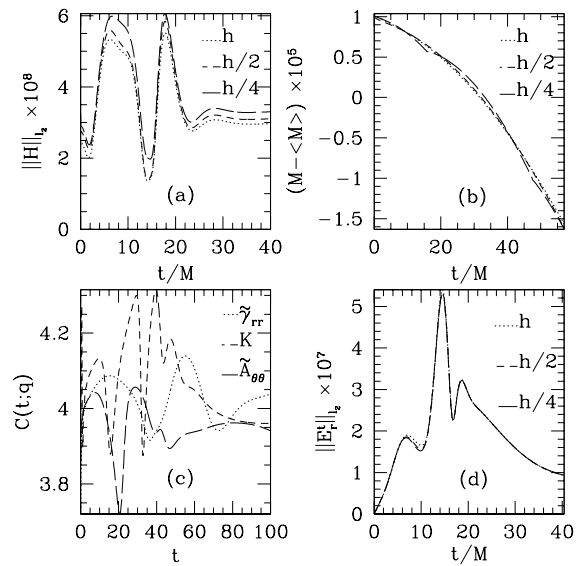}
		\caption[Testing the Numerical Solver]{\label{fig:tests}
			Results from various tests that verify the
			accuracy and consistency of our numerical solver and the finite
			differencing method used to integrate the equations.
         (a)~The evolution of the $l_2$-norm (RMS value) of the Hamiltonian constraint.
	      The norm is plotted for 3 different resolutions $h$,
         $h/2$ and $h/4$ corresponding to $N_r=512$, $1024$ and $2048$, respectively.
			The data for the $N_r=1024$ and $N_r=2048$ curves have been rescaled by factors of 
         $4$ and $16$, respectively, and the overlap of the three lines thus signals the expected second
			order convergence to zero of the constraint deviation.
         We observe similar convergence properties for
         the momentum constraint as well as the constraint 
			equation (\ref{eq:deflam}) for $\tilde{\Lambda}^i$, 
			and the constraint~(\ref{eq:tracea})
         for the trace of $\tilde{A}_{ij}$.
         Additionally, since the operator used to evaluate the residual 
			of the Hamiltonian constraint
         is distinct from that used in the determination of 
			the initial data, the test also validates 
         the initial data solver.
			(b)~Conservation of the ADM mass during the evolution 
			of strong initial data. Here the deviation of the mass
			from its time average is plotted for 3 different resolutions. 
			Higher resolution values have again been rescaled so that overlap
         of the curves demonstrates $O(h^2)$ convergence to 0
         of the deviation of the total mass.
			(c)~The convergence factor defined in (\ref{eq:convf})
 		   for three of the primary BSSN variables: $\tilde{\gamma}_{rr}$,
			$K$, and $\tilde{A}_{\theta\theta}$.  In the limit $h\to0$ we expect all
         curves to tend to the constant 4. The plot thus provides 
         evidence for second order convergence of all of the values throughout
         the evolution.  All of the other primary BSSN variables as well as the dynamical 
         scalar field quantities demonstrate the same convergence. 
			(d)~Direct verification that the metric 
			found by numerically solving the BSSN equations satisfies 
			Einstein's equations in their covariant form. 
			Here the $tr$ component of the 
			residual $E^{\mu}_{\;\nu}$ defined in (\ref{eq:rese}) is
			plotted for 3 different resolutions. 
		   Once more, higher resolution
         values have been rescaled so that overlap of the curves signals the expected $O(h^2)$ 
         convergence of the residuals to 0.  
			All of the plots correspond to evolution of strong subcritical 
			initial data with 1+log slicing.
         For (a) and (b) the shift vector was set to 0, while in (c) and (d) it
         was evolved using the Gamma-driver condition (i.e.~$\beta=\betagamma$).
			}
	\end{center}
\end{figure}

\subsection{Tests \label{sec:tests}}

%
%
%
%
%
%

This section documents various tests we have made to validate the correctness of
our numerical solver as well as the consistency of the finite-differencing
method used to evolve the system of G-BSSN equations. We use a variety of diagnostic tools, including
monitoring of the constraint equations, convergence tests of the primary
dynamical variables, and a direct computation to check if the 
metric and matter fields calculated via the G-BSSN formulation satisfy the covariant form of 
Einstein's equations.
All of the calculations were performed using the 
1+log slicing condition (\ref{eq:1pl}) and either $\beta=0$ or $\beta=\betagamma$ where
$\betagamma$ satisfies
the Gamma-driver condition~(\ref{eq:betagammadrive}).

\subsubsection{Constraints and conserved quantities\label{sec:constraints_conserved}}
We monitor the evolution of the constraint equations (\ref{eq:HC},\ref{eq:MC})
during a strongly-gravitating evolution where the nonlinearities of
the equations are most pronounced. 
As demonstrated in Fig.~\ref{fig:tests} (a),
at resolutions typical of those used in our study, the Hamiltonian constraint is well preserved during 
such an evolution and, importantly, 
the deviations from conservation converge to zero at second order in the mesh spacing as expected.

The total mass-content of the spacetime seen at spatial
infinity (the ADM mass) is a conserved quantity.
Here, using the G-BSSN variables the Misner-Sharp mass function is given by
\begin{equation}
M(r) = \frac{r \B^{1/2}e^{2\phi} }{2}\left[ 1 -\frac{\B}{\A}\left(1 + r\frac{\partial_r \B}{2\B} + 2r\frac{\partial_r e^\phi}{e^\phi}  \right)^2 \right] \, .
\end{equation}
The total mass, $M_{\mathrm{ADM}}$, can be evaluated at the outer boundary,
\begin{equation}
M_{\mathrm{ADM}} \equiv M(r=\infty) \, .
\end{equation}
The deviation of the total mass from its time average 
is plotted in Fig.~\ref{fig:tests}(b); as
the resolution of the numerical grid increases the variations converge to zero
in a second order fashion.

\subsubsection{Convergence test}
As mentioned in Sec.~\ref{sec:initialization} and Sec.~\ref{sec:evolution},
we implemented our code using second order finite differencing of all spatial and temporal derivatives.
Denoting any continuum solution component by $q(t,X)$, where $X$ is the spatial coordinate,  
and a discrete approximation to it computed 
at finite difference resolution, $h$, by $q^h(t,X)$, to leading order in $h$ we expect
\begin{equation}
   q^h(t,X) = q(t,X) + h^2 e_2[q](t,X) + \ldots \, .
   \label{eq:richardson}
\end{equation}
Fixing initial data, we perform a sequence of calculations with resolutions $h$, $h/2$ and $h/4$
and then compute a convergence factor, $C(t;q)$, defined by
\begin{equation}
C(t;q) = \frac{||q^h(t,X)-q^{h/2}(t,X)||_2}{||q^{h/2}(t,X)-q^{h/4}(t,X)||_2} \, ,
\label{eq:convf}
\end{equation}
where $||\cdot||_2$ is the $l_2$ norm, i.e.~the root mean square (RMS) value.
It is straightforward to argue from~(\ref{eq:richardson}) that, for sufficiently small $h$,
$C(t,q)$ should approach 4 if the solution {\em is} converging at second order.
The values of 
the convergence factor for a selection of dynamical variables are plotted for a strong-data evolution 
in Fig.~\ref{fig:tests}(c) and provide clear evidence that the solution
is second-order convergent throughout the time evolution.

\subsubsection{Direct validation via Einstein's equations}
A direct method to test the fidelity of our numerical solver involves
the evaluation of a residual based on the {\em covariant} form of Einstein's equations.
%
We start with 
a reconstruction of the four-dimensional metric in spherical symmetry,
\begin{equation}
	ds^2 = (-\alpha^2+\beta^2a^2)dt^2 + 2a^2\beta dtdr + a^2dr^2 + r^2b^2d\Omega^2 \, ,
	\label{eq:metricfullspher}
\end{equation}
using the primary G-BSSN variables, $\tilde{\gamma}_{ij}$ and $\phi$.
In particular, $a$ and $b$ are simply given by
\be
a(t,r) = e^{4\phi(t,r)}\A(t,r) \, ,
\ee
\be
b(t,r) = e^{4\phi(t,r)}\B(t,r) \, .
\ee
We then check to see if the metric~(\ref{eq:metricfullspher})  satisfies the 
covariant Einstein
equations~(\ref{eq:einsteinequatoin}) to the expected level of truncation error. 
Specifically, defining
the residual
\begin{equation}
	E^{\mu}_{\;\nu} \equiv G^{\mu}_{\;\nu} -8\pi T^{\mu}_{\;\nu} \, ,
	\label{eq:rese}
\end{equation}
and replacing all derivatives in $G^\mu_\nu$ with second order finite differences,
we expect $E^{\mu}_{\;\nu}$ to converge to zero as $O(h^2)$ as $h\to0$.\footnote{Although
it is not crucial for the usefulness of this test, we discretize the $E^\mu_\nu$ using 
a difference scheme that is distinct from the one used in the main code.}
Precisely this 
behaviour is shown in Fig.~\ref{fig:tests}(d). 
This is a particularly robust test of our implementation 
since the non-trivial components of the covariant Einstein equations
are quite complicated and, superficially at least, algebraically independent of 
the BSSN equations.
For instance, the $tr$ component of the residual (\ref{eq:rese}) is given by
\begin{eqnarray}
	E^t_{\;r} &=&\frac{2\beta}{r\alpha^2}\left( \xf{a} - 2\xf{b} + \xf{\alpha}  \right) \nonumber \\
&+&\frac{2\beta}{\alpha^2}\left( -\frac{\partial_r^2 b}{b} +\frac{\partial_r a \partial_r b}{ab} - \frac{(\partial_r\Psi)^2}{2} + \frac{\partial_r \alpha \partial_r b}{\alpha b}    \right) \nonumber \\
&+&\frac{2}{\alpha^2}\left( \frac{\partial_t \partial_r b}{b} + \frac{\partial_r \Psi \partial_t \Psi}{2} -\frac{\partial_t a \partial_r b}{ab} - \frac{\partial_t b \partial_r\alpha}{\alpha b} \right)\nonumber \\
&+&\frac{2}{r\alpha^2}\left( -\tf{a} + \tf{b} \right) \, 
	\label{eq:etr}
\end{eqnarray}
and depends on all of the dynamical variables of the system.
The observed convergence of the residual
is only plausible if (1) our G-BSSN equations~(\ref{eq:evolphi}-\ref{eq:evolgamma}) have 
been correctly derived from the covariant Einstein
equations, (2) we have discretized the geometric and matter equations properly, and 
(3) we have solved the full set of discretized equations correctly. 

\subsection{Finding black hole threshold solutions \label{sec:fbh} }
The strength of the initial data can be set by adjusting the 
amplitude of the scalar field, $p$, in (\ref{eq:initpsi}).
For weak enough initial data (small enough $p$), the matter shell will
reach the origin and then disperse, with the final state 
being a flat spacetime geometry. Sufficiently strong initial data, on the other 
hand (large enough $p$), results in a matter concentration in the vicinity 
of the origin which is sufficiently self-gravitating that a black hole forms.
Using a binary search, we
can find the threshold initial data,
defined by $p=p^\star$, for which $p<p^\star$ results in dispersal
while $p>p^\star$ yields black hole formation.  At any stage of the 
calculation, the binary search is defined by two 
``bracketing'' values, $p_l$ and $p_h$, such that evolutions with $p=p_l$ and 
$p=p_h$ result in dispersal and black hole formation, respectively.  It 
is convenient to define the amount of parameter tuning that has 
occurred by the dimensionless quantity
\be
   \delta p \equiv \frac{p_h - p_l}{p_l} \, .
   \label{eq:deltap}
\ee

The dispersal case can be detected easily as the scalar field leaves the vicinity
of the origin and the geometry approaches flat spacetime. 
To detect black hole formation, we use an apparent horizon finder
to locate a surface $r={\rm const.}$ on which the 
divergence of the outgoing null rays 
vanishes. We first define the divergence function
\begin{equation}
	\Theta = q^{\mu\nu} \nabla_\mu k_\nu \, ,
	\label{eq:Theta}
\end{equation}
where $q^{\mu\nu}$ is the induced metric on the constant $r$ surface. In spherical
symmetry with metric (\ref{eq:metricfullspher}) we have
\begin{equation}
	q_{\mu\nu} = \mathrm{diag}\left( 0, 0 , r^2b^2,r^2b^2\sin^2\theta \right) \, ,
\end{equation}
where $k^\mu$ is the null outgoing vector given by
\begin{equation}
	k_\mu = \frac{1}{\sqrt{2}}\left[a\beta-\alpha,a,0,0\right] \, .
\end{equation}
Therefore, (\ref{eq:Theta}) becomes
\begin{equation}
	\Theta = \frac{\sqrt{2}}{rb}\left(\frac{r}{\alpha}\partial_t(b) +\left( \frac{1}{a} - \frac{\beta}{\alpha} \right)\partial_r(rb)   \right) \, .
\end{equation}
The formation of an apparent horizon\footnote{Technically a marginally trapped surface---the apparent horizon being the outermost of these.} is signaled by 
the value of the function $\Theta$ crossing zero at some radius and, modulo
cosmic censorship, implies that the spacetime contains a black hole.  We note 
that since the focus of our work was on the critical (threshold) solution 
we made no effort to continue evolutions beyond the detection of trapped
surfaces.

\section{Results \label{sec:results}}

In this section we describe results from two sets of numerical experiments to
study the efficacy of the G-BSSN formulation in the context 
of critical collapse.  Again, our calculations use 
the standard 1+log slicing condition for the lapse, and a shift which is either zero or determined from the 
Gamma-driver condition.

\begin{figure}
	\begin{center}
		\includegraphics[width=0.44\textwidth]{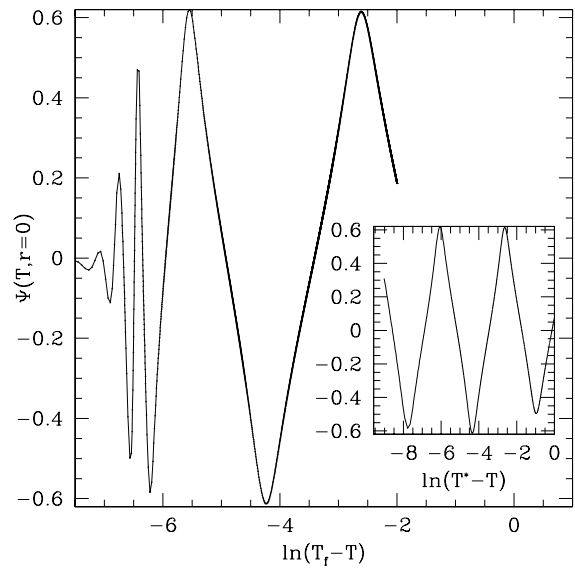}
		\caption[Echoing of Scalar Field]{  \label{fig:phi0}
	    Echoing behaviour in the scalar field for a marginally subcritical 
       evolution with $\delta p \approx 10^{-12}$.  
       The main plot displays the central value of the scalar 
		 field versus a logarithmically scaled time 
       parameter, $\ln(T_f - T)$, where $T$ is central proper 
		 time and $T_f$ is the approximate
       value of that time when near-critical 
		 evolution ceases and the total dispersal of the pulse
       to infinity begins. 
       This particular scaling 
		 is chosen solely to more clearly demonstrate the evolution
       of the central value of $\Psi$ during the critical phase through
       to dispersal.  Note that our choice of abscissa means that evolution
       proceeds from right to left. 
       The inset also plots $\Psi$ at $r=0$ but 
		 now in the ``natural'' logarithmic time coordinate 
		 $\tau\equiv \ln (T^\star - T)$ where $T^\star$ is 
		 the ``accumulation time'' at which the 
       solution becomes singular and which has been 
		 estimated based on the positions of the extrema in $\Psi$. 
		 The amplitude of the scalar field at the origin oscillates 
		 between $(-0.61,0.61)$, consistent 
		 with the calculations reported in~\cite{Choptuik:1992}.
       The data yield an echoing exponent
		 of $\Delta = 3.43\pm 0.02$  which is in agreement with
       the value $\Delta = 3.445452402(3)$ 
		 Martin-Garcia and Gundlach  have computed 
       by treating the computation of the precisely-critical solution
       as an eigenvalue problem 
       \cite{MartinGarcia:2003gz}.
	 }
	\end{center}
\end{figure}

\def\R{{\cal R}}
\def\Rmax{{\cal R}_{\rm max}}
\begin{figure}
	\begin{center}
		\includegraphics[width=0.44\textwidth]{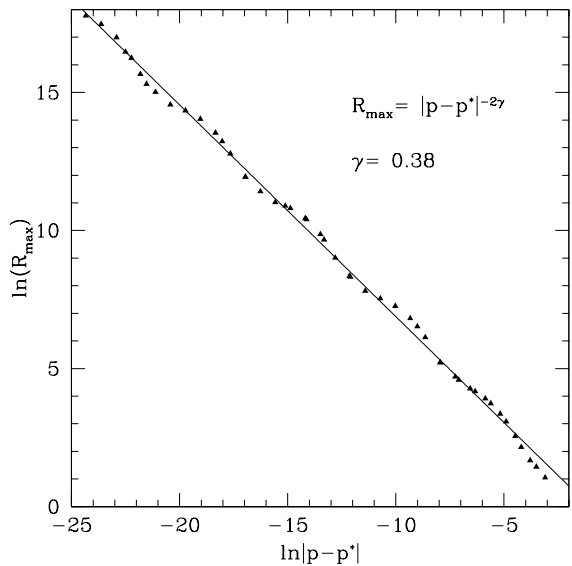}
		\caption[Scaling of Ricci Scalar]{  \label{fig:scalebeta}
        The maximum central value, $\Rmax$, of the four-dimensional Ricci scalar, $\R$, attained during 
		  subcritical evolution as a function of the logarithmic 
        distance 
        $\ln|p-p^\star|$
        of the tuning
		  parameter from the critical value.
        As first observed in 
		  \cite{Garfinkle:1998} 
        the Ricci
		  scalar scales as $R\sim|p-p^\star|^{-2\gamma}$,
        where $\gamma$ is
		  the universal mass-scaling exponent in (\ref{eq:massscale}). 
        The value $\gamma = 0.38 \pm 0.01$ computed via a least squares fit is in agreement with 
		  the original calculations \cite{Choptuik:1992} 
		  as well as many other subsequent computations.  
		  We note that the oscillations of the data about the linear fit 
        are almost certainly genuine, at least in part.  As discussed in the text, we {\em expect}
        a periodic wiggle in the data with period $\Delta/(2\gamma) \approx 4.61$. Performing 
        a Fourier analysis of the residuals to the linear fit we find a peak at about 4 with a bandwidth
        of approximately 1, consistent with that expectation.
        As described in more detail in the text, although we have data from computations with  
        $\ln|p-p^\star| < -25$, we do not include it in the fit.  The naive method we use to 
        estimate $p^\star$ means that the relative uncertainty 
		  in $p-p^\star$ grows substantially 
        as $p\to p^\star$ so that inclusion of the data 
		  from the most nearly-critical calculations 
        will corrupt the overall fit.
    }
	\end{center}
\end{figure}

\subsection{Zero shift}



We first perform a collection of numerical experiments where 
the shift vector is set to zero. 
As described in Sec.~\ref{sec:fbh}, in principle we can 
find the
black hole threshold solution $p\simeq p^\star$ using a binary search 
algorithm which at any stage is defined by two values $p_l$ and $p_h$, with 
$p_l < p < p_h$, and where $p_l$ corresponds to dispersal (weak data) while 
$p_h$ corresponds to black hole formation (strong data).

As discussed in the introduction,
the massless scalar collapse model has a very well-known 
critical solution, and we summarize the features most relevant to our study here.  
The threshold configuration is discretely self-similar 
with an echoing exponent measured from the first calculations to be 
$\Delta \approx  3.44$ \cite{Choptuik:1992}. 
%
%
Following the original studies, Gundlach \cite{Gundlach:1995}
showed that the construction of the precisely discretely self-similar spacetime could 
be posed as an eigenvalue problem, the solution of which led to the 
more accurate value $\Delta=3.4439 \pm 0.0004$. This estimate was subsequently improved by Martin-Garcia and
Gundlach
to 
$\Delta = 3.445452402(3)$~\cite{MartinGarcia:2003gz}.

The original calculations determined a value $\gamma \approx 0.37$ for the mass-scaling 
exponent~\cite{Choptuik:1992};
further work based on perturbation theory gave 
$\gamma \approx 0.374$~\cite{PhysRevD.55.695, PhysRevD.58.064031}.
Here it is important
to note that, as pointed out independently by Gundlach~\cite{PhysRevD.55.695} and 
Hod and Piran~\cite{PhysRevD.55.R440}, the 
simple power law scaling~(\ref{eq:massscale}) 
gets modified for discretely self-similar critical solutions to
\begin{equation}
   \ln M = \gamma \ln \left\vert p-p^\star\right\vert + c + f\left(\gamma\ln\left\vert p-p^\star\right\vert+c\right) \, ,
\end{equation}
where $f$ is a universal function with period $\Delta$ and $c$ is a constant depending on the initial 
data.  This results in the superposition of a periodic ``wiggle'' in the otherwise linear
scaling of $\ln M$ as a function of $\ln | p - p^\star|$.

Finally, Garfinkle and Duncan~\cite{Garfinkle:1998} pointed out that near-critical scaling is seen in physical quantities other
than the mass and, dependent on the quantity, in the subcritical as well as supercritical regime.  
In particular they argued that in subcritical evolutions the maximum central value, $\Rmax$, 
of the four-dimensional
Ricci scalar, $\R$, defined by
\begin{equation}
   \Rmax \equiv \max_t {\cal R}(0,t) \, ,
\end{equation}
should satisfy the scaling
\begin{equation}
   \Rmax \sim |p-p^\star|^{-2\gamma} \, ,
   \label{eq:rmaxscaling}
\end{equation}
where the factor $-2$ in the scaling exponent can be deduced from the fact that the curvature 
has units of ${\rm length}^{-2}$.  For the discretely self-similar case this scaling law is also modulated by
a wiggle with period
$\Delta/(2\gamma)$,
which for the massless scalar field is about 4.61.
\begin{figure}
	\begin{center}
		\includegraphics[width=0.44\textwidth]{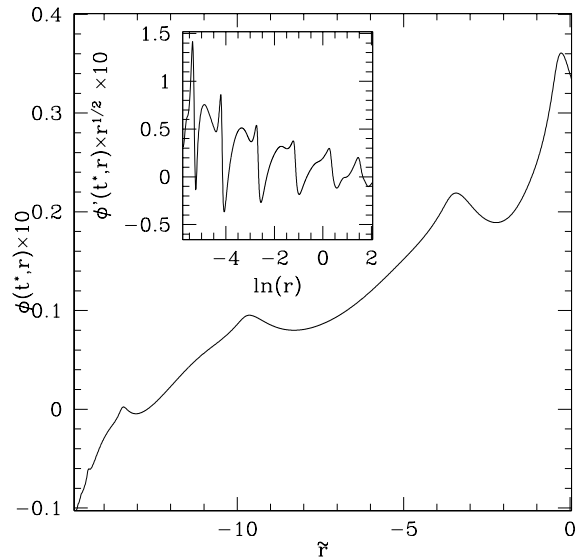}
		\caption[Echoing in Geometry]{  \label{fig:phi}
	Discrete self-similarity of the geometry of
	spacetime in the black hole threshold evolution previously 
   discussed in Fig.~\ref{fig:phi0}. Here, the
    G-BSSN variable $\phi$ is plotted as a function of the computational
    radial coordinate
    $\tilde{r}$ at the accumulation time $t^\star$. 
	 Note that from (\ref{eq:phiratio}) $\phi$ measures the
	 deviation of the determinant of the 3-metric from that of a flat metric.    
	 The inset graph is the radial derivative of $\phi$
	scaled by $\sqrt{r}$ to
   highlight the formation of fine structure in the geometry
	of the critical solution.  
   The approximate periodicity of $\sqrt{r} \phi'$ in $\ln(r)$ (modulo an overall
   varying scale)
   provides weak evidence that the coordinate system used in the calculation
   adapts to the self-similarity of the critical solution.
}
	\end{center}
\end{figure}

\begin{figure}
	\begin{center}
		\includegraphics[width=0.49\textwidth]{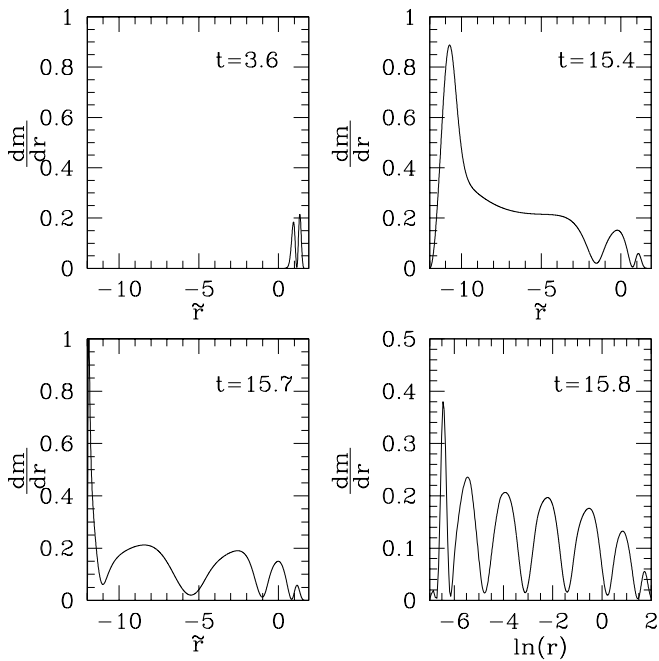}
		\caption[Mass Density Function in Critical Regime]{  \label{fig:dmdr}
		 Snapshots of radial mass density for a marginally subcritical calculation 
       ($\delta p \approx 10^{-12}$, $N_r=2048$). 
       Plotted is $dm/dr=r^2\rho(t,r)$
		 where $\rho$ is defined by~(\ref{eq:defrho}). 
       In this calculation $\beta=0$ so we also have
       $dm/dr = T^t_{\;t}$.
		 As the solution evolves,
       development of echos is clearly seen.  In the final frame, which is 
       at an instant $t=15.8$ that is close to the accumulation time $t^\star$,
       we observe 4 echos. Note that we do not count the tall thin peak at the extreme 
       left nor the first two peaks on the right as echos.  The skinny 
       peak will develop into an echo as $p$ is tuned closer to $p^\star$. 
       The two peaks on the right
       account for the bulk of the matter and represent the part of the initial pulse that
       implodes through the origin and then disperses ``promptly'', i.e.~without
       participating in the strongly self-gravitating dynamics.  A corresponding 
       plot for an evolution far from criticality would contain {\em only} those 
       two peaks.
       Note that the first three plots 
		 use the computational coordinate $\tilde{r}$ to provide a sense of
       the actual numerical calculation,
		 while the last plot uses $\ln(r)$ in order to best highlight
		 the discrete self-similarity of the threshold solution.  
       As is the case for the data plotted in the inset of the previous figure, the 
       approximate periodicity of the mass density in $\ln(r)$ suggests 
       that the coordinates are adapting to the self-symmetry of the critical
       spacetime.
		}
	\end{center}
\end{figure}

%
Using initial data given by~(\ref{eq:initpsi}) we tune $p$ so that it is 
close to the critical value:  typically this involves reducing the value of $\delta p$ 
defined by~(\ref{eq:deltap}) so that it is about $10^{-12}$, which is a few
orders of magnitude larger than machine precision.
Our implementation includes code that actively monitors the dynamical variables for any indications 
of coordinate singularities or other pathologies which could
cause the numerical solver to fail.   Provided that such pathologies do not 
develop, we expect to observe features characteristic of critical 
collapse---discrete self-similarity and mass scaling in particular---to emerge as $p\to p^\star$.

One way the discrete self-similarity of the critical solution is manifested is as a sequence of 
``echoes''---oscillations of the scalar field
near the origin such that after each oscillation the profile of the scalar field is 
repeated but on a scale $\exp(\Delta)$ smaller than that of the preceding echo 
(see Eq.~(\ref{eq:echoing})).
The oscillations are similarly 
periodic in the logarithmic
time scale $\ln(T^\star-T)$, where 
$T$ is the proper time measured at the origin,
\be
T(t) \equiv \int_{0}^{t} \alpha(\hat{t},0)\,d\hat{t} \, ,
\ee
and $T^\star$ is the accumulation 
time at which the singularity forms (always at $r=0$).
Furthermore, viewed at the origin, the oscillations of the scalar field 
occur at a fixed amplitude of about $0.61$ (with our units and conventions for the 
Einstein's equations). As shown in Fig.~\ref{fig:phi0},
when we tune the initial data to the critical value,
the central value of scalar field exhibits oscillatory behaviour
and the amplitude
is close to the expected value. The anticipated periodicity in logarithmic time
is also apparent with a measured 
$\Delta=3.43\pm 0.02$, in agreement with previous results. We thus have 
strong evidence that the evolution has indeed 
approached the critical regime
and that the measured oscillations are true echos rather than numerical artifacts.

Evidence that our code correctly captures the expected critical scaling 
behaviour~(\ref{eq:rmaxscaling})
of $\Rmax$ 
is presented in 
Fig.~\ref{fig:scalebeta}. 
We find $\gamma = 0.38 \pm 0.01$, consistent with previous calculations.
%
%
We note that we can measure scaling from our computations 
up to $\ln|p-p^\star| = -29$ (or $|p-p^\star|\approx10^{-13}$).
However, in Fig.~\ref{fig:scalebeta} we have excluded the last few values closest
to the critical point from both the plot and the linear fit: specifically, 
we truncate the fit at 
$\ln|p-p^\star| = -25$.
The rationale for this is that 
we use the largest subcritical value of $p$ 
as an approximation to the critical value $p^\star$ rather than, for example,
implementing a multi-parameter fit that includes $p^\star$ as one of the 
parameters. Our estimate of $p^\star$ thus has an
intrinsic error of $e^{-29} \approx 10^{-13}$ and 
by fitting to data with $\ln|p-p^\star |\ge -25$ we render the  error 
in the $p^\star$ estimate
essentially irrelevant. 
We note that consistent with the early observations of the robustness of mass scaling
in the model~\cite{Choptuik:1992}, measuring the exponent $\gamma$ can be achieved by moderate
tuning, in this case $\ln|p-p^\star|\approx -9$, (i.e.~$\delta p \approx 10^{-3}$). 
However, to be able to
observe the echoing exponent (the oscillations around the fitting line, for example)
we need to tune much closer to the critical value. 

The echoing behaviour of the critical solution is also reflected in the geometry of spacetime and
the matter variables other than the scalar field.
Fig.~\ref{fig:phi} shows the radial profile of the G-BSSN variable 
$\phi$ at an instant close to the accumulation time $T^\star$. As seen in
this plot, fine structure develops in the function in the near-critical regime.
Observe that from the definition (\ref{eq:decomgam}) and the choice (\ref{eq:lagchoice}), the
scalar $\phi$ is the ratio of the determinant of the 3-metric, $\gamma$, to
the determinant of the flat metric, $\mathring{\gamma}$:
\be
\phi = \frac{1}{12}\ln(\gamma/\mathring{\gamma}) \, . 
\label{eq:phiratio}
\ee

The radial matter density, $dm/dr = r^2\rho$,
is a convenient diagnostic quantity for viewing near-critical evolution. 
Snapshots of this function from a typical marginally subcritical calculation
are shown in Fig.~\ref{fig:dmdr}: the echoing behavior is clearly evident in 
the sequence.  The number of echos is dependent on the degree to which the
solution has been tuned to criticality.  In this case, where $\delta p = 10^{-12}$,
we expect and see about 4 echos (last frame of the figure).
Here we note that
each of the echos in $dm/dr$ corresponds to half of one of the scalar field oscillations
shown in Fig.~\ref{fig:phi0} (where the inset shows about $2\frac12$ full cycles).

\begin{figure}
	\begin{center}
		\includegraphics[width=0.49\textwidth]{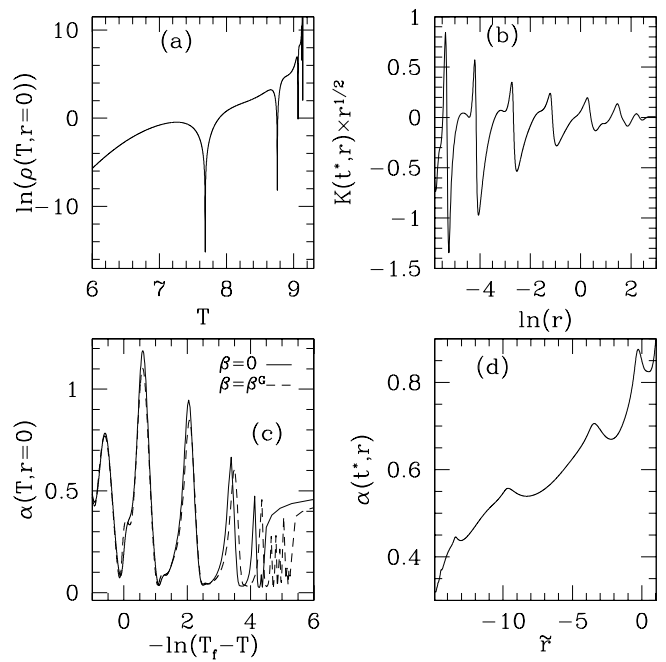}
		\caption[Matter and Geometry at Critical Point]{  \label{fig:crit}
		 Profiles of matter and geometry variables from 
		 strongly gravitating, near-critical evolutions where the echoing behaviour
       emerges.  Results were computed using 1+log slicing and zero shift, except
      for the dashed curve in (c) where $\beta=\beta^G$.
		 (a) Central energy density, $\rho(T,0)$, as a function of proper
		 central time, $T$,
		 and in logarithmic scale for a supercritical evolution. 
		 The density oscillates and grows exponentially
		 as the system approaches the critical
		 solution and then eventually collapses to form a black hole. (b)
		 Profile of the extrinsic curvature, $K(t^{\star},r)$---scaled by $r^{1/2}$ in order to 
       make the echoing behaviour more visible---where $t^\star$ 
		 denotes a time very close to the accumulation time.  
		 The evolution is marginally subcritical
       in this case.
       (c) Central value of the lapse function, $\alpha$,
       during subcritical evolutions with $\beta=0$ (solid)
       and $\beta=\beta_G$ (dashed). The plots use a logarithmically 
       transformed proper time variable, $-\ln(T_f-T)$, where $T_f$ is the 
       approximate time at which the final dispersal of the pulse from 
       the origin begins.
		 In both cases $\alpha$ exhibits echoing
		 and there is no evidence of pathological 
       behaviour, such as the lapse collapsing or becoming negative.  The close 
       agreement of $\alpha$ for the two choices of $\beta$ indicates 
       that the time slicing varies little between the two coordinate systems.
       Note that there are three extra oscillations for the $\beta=\beta_G$ case,
       in the time interval $-\ln(T_f-T) \gtrsim 4.5$.
       These are spurious and due to a lack of finite-difference resolution; there
       are only 6 time steps in each oscillation.
       (d)~Radial profile $\alpha(t^\star,r)$ at a 
		 time $t=t^\star$ which is close to the accumulation time
       and when the self-similarity and echoing in the spacetime geometry is apparent.
		}
	\end{center}
\end{figure}

Fig.~\ref{fig:crit}(a) plots the central matter density $\rho(t,0)$ for a marginally
supercritical calculation.
In accord with the self-similar nature of the near-critical solution,
the central density grows exponentially with time.
Fig.~\ref{fig:crit}(b) is a snapshot of the extrinsic curvature at the
critical time $t\approx t^\star$ while Fig.~\ref{fig:crit}(c) 
shows the dynamics of the central value of the lapse function and compares it with 
$\alpha$ from the calculations performed with $\beta=\beta_G$ described in the 
next section.
Fig.~\ref{fig:crit}(d) displays the profile of the lapse at the critical time.

\begin{figure}
	\begin{center}
		\includegraphics[width=0.49\textwidth]{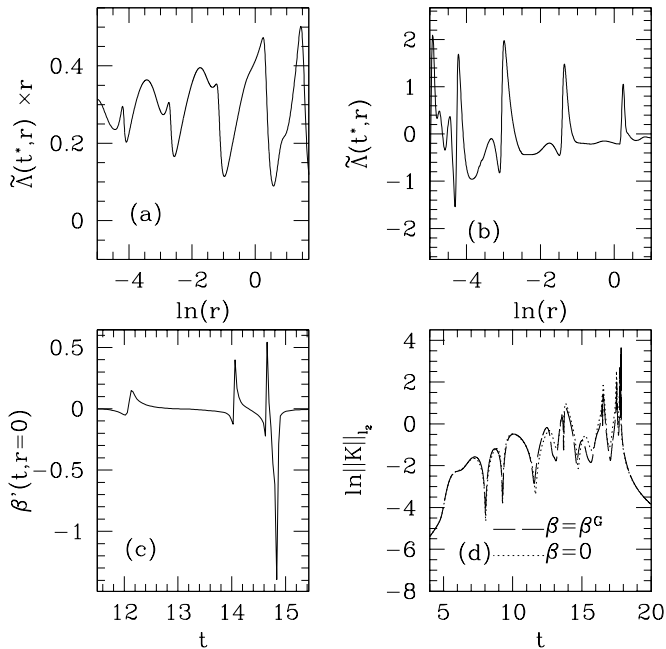}
		\caption[Evolution of G-BSSN Variables in Critical Regime]{  \label{fig:coord}
         Profiles of various G-BSSN variables from marginally subcritical evolutions.
       (a)~Profile of the conformal connection $\tilde{\Lambda}$ as computed with $\beta=0$
       and at a time $t^\star$ close to the accumulation time.
		 Note that the function has been scaled 
		 by $r$ and in fact diverges like $1/r$. 
		 (b)~Profile of $\tilde{\Lambda}$ as computed with $\beta=\beta^G$, again at a moment
       close to the accumulation time.
       Here the $1/r$ growth seen when the condition $\beta=0$ is adopted is absent.
		 (c)~Profile of the central spatial derivative of the 
		 shift vector, $\beta'(t,0)$, as computed with $\beta=\beta^G$. 
       As the echos develop closer to the origin, $\beta'$ 
       increases and presumably will diverge in the continuum, precisely-critical limit.
       (d)~Time development of the $l_2$-norm of the extrinsic curvature during
       subcritical evolutions for both the $\beta=0$ and $\beta=\beta^G$ calculations.
       In both cases the extrinsic curvature
       develops a divergent profile near $r=0$ in the critical regime.
	}
	\end{center}
\end{figure}


We note that we have not fully resolved the 
critical behaviour in the sense of having tuned $p$ to 
the limit of machine precision, $\delta p\approx 10^{-16}$, 
which would capture roughly 2 additional 
echos in the threshold solution 
(one full echo in the scalar field).   
%
In principle, by setting $N_r$ sufficiently large 
we could almost certainly do so since there are no indications that our 
method would break down at higher resolution and closer 
to criticality.   However, we estimate that the required compute 
time for a complete critical search would increase 
from weeks to several months and we have thus not done this.  
Ultimately, a more effective approach to enhancing the resolution would be to 
incorporate a technique such as adaptive mesh refinement into our solver.

The results displayed in Figs.~2--6 provide strong evidence that the coordinate system consisting
of 1+log lapse and zero shift remains non-singular in the critical regime, at least for the range of 
scales probed for $\delta p \approx 10^{-12}$.  Additionally, the approximate periodicity in $\ln(r)$ 
that can be seen, for example, in $\sqrt{r}\phi'$ (Fig.~4) and $dm/dr$ (Fig.~5) suggests that the coordinates
may be adapting to the self-similarity.  Whether or not this is actually the case is a matter requiring
further study.
\subsection{Gamma-driver shift}
We now briefly report on experiments similar to those of the previous section but where 
the shift was evolved with the
Gamma-driver condition (\ref{eq:gammadrive}).
A principal observation is that this gauge 
also facilitates near-critical evolutions 
with results similar to the $\beta=0$ choice.  In particular, we are again able
to observe all of the characteristics
of the black hole threshold solution.

The gauge condition (\ref{eq:gammadrive}) acts as a damping factor for 
the conformal connection, $\tilde{\Lambda}^i$, and we would therefore 
expect to observe a significant change in the
profile of $\tilde{\Lambda}^i$ at threshold relative to the zero-shift case. 
This expectation is borne out by the comparison
illustrated in Figs.~\ref{fig:coord}~(a) and (b). When $\beta=0$, $\tilde{\Lambda}^i$ diverges 
as $1/r$ close to the origin while it appears to have finite amplitude 
for $\beta=\betagamma$.
We find that the shift develops very sharp oscillations near the origin; some typical behaviour 
can be seen in the plot of $\beta'(t,0)$ shown in~ Fig.~\ref{fig:coord}(c).
We believe that these oscillations are genuine and our expectation is that $\beta'(t,0)$ will 
diverge in the precise critical limit.
Further, 
we observe that the oscillations can create numerical artifacts and generally require
higher resolution relative to the $\beta=0$ case, as well as dissipation, to be controlled.
Indeed, 
when using the Gamma-driver condition
we find that 
Kreiss/Oliger dissipation is crucial 
to suppress
unresolved high frequency oscillations close to the origin.
Fig.~\ref{fig:coord}~(d)
shows the growth in the norm of the extrinsic curvature during a subcritical
evolution. The norm of $K$ does not exhibit any significant difference for 
the two choices of the shift.

As was the case for the $\beta=0$ calculations, the results shown in Figs.~6 and 7
strongly suggest that the combination of 1+log slicing and Gamma driver shift
provides a coordinate system which is adequate for computing the near-critical 
solution.  In addition, the approximate periodicity seen in Figs.~6(b), 6(c),
7(a) and 7(b) suggest that this gauge may also be adapting to the self-symmetry.

\section{Conclusion \label{sec:conclusion}}
We have described a numerical code that implements
a generalized BSSN formulation adapted to spherical symmetry. 
Using standard dynamical coordinate choices, including 1+log slicing and a shift
which either vanished or satisfied a Gamma-driver condition, we
focused specifically on the applicability
of the formulation and the gauge choices to studies of type II critical phenomena.
As a test of the approach we revisited the model of massless scalar collapse, where 
the properties of the critical solution are very well known from previous work.
For both choices of the shift, we found that our code was able to generate evolutions that were very 
close to criticality so that, in particular, we could observe the 
expected discrete self-similarity of the critical solution.  
To our knowledge, this is the first fully evolutionary implementation of a
hyperbolic formulation of Einstein's equations that has been able to unequivocally resolve 
discrete self-similarity in type II collapse.
Furthermore, 
measured properties from near-critical solutions, including the mass-scaling and 
echoing exponents, are in agreement with previous work.
Our results strongly suggest that the G-BSSN formulation, in conjunction
with standard dynamical coordinate conditions, is capable of evolving the spacetime
near criticality without the development of coordinate pathologies.
There is also some evidence that both gauges adapt to the self-similarity, but we 
have not yet studied this issue in any detail.

We found that certain of the primary G-BSSN variables diverge 
as the critical solution is approached: this is only to be expected since the precisely
critical solution contains a naked singularity.  
Dealing with such solution
features in a stable and accurate manner presents a challenge for any code and in our case we found 
that a combination of a non-uniform grid and Kreiss/Oliger dissipation was crucial.
Our use of a time-implicit evolution scheme may have also been important 
although we did not experiment with that aspect of our implementation.  However, we suspect 
that the implicit time-stepping helped maintain regularity of the solutions near $r=0$,
as other researchers have found.



Given the success of the G-BSSN approach, it is natural to consider its 
generalization and application
to settings with less symmetry, but where curvilinear coordinates are still adopted.
In particular, one axisymmetric problem  that has yet to be resolved
is the collapse of pure gravity waves.
This scenario
arguably provides the most fundamental critical phenomena 
in gravity as the behaviour must be intrinsic to the Einstein equations, rather than being
dependent on some matter source.
Critical collapse of gravitational waves--with mass scaling and echoing---was
observed by Abrahams and Evans over 20 years ago~\cite{Abrahams:1993}.
However, their original results have proven very difficult to reproduce (or refute)
\cite{Alcubierre:1999, Garfinkle:2000, Rinne:2008, Sorkin:2010, Hilditch:2013}. 
We refer the reader to the recent paper by Hilditch 
et~al.~\cite{Hilditch:2013} for
detailed discussions concerning some apparent inconsistencies among the follow-up studies, as well as
the challenges and complications involved in evolving various types of nonlinear gravitational
waves.  We are currently extending the methodology described above to the 
axisymmetric case with plans to use the resulting code to study vacuum collapse. 
Results from this undertaking will be reported in a future paper.




\begin{acknowledgments}
This research was supported by NSERC, CIFAR and by a 4FY 
scholarship to AA from UBC. The authors thank William G. Unruh for 
insightful comments and a thorough reading of a draft of this paper.
\end{acknowledgments}

\def\Beta{\mathscr{B}}
\appendix
\section{BSSN in Spherical Symmetry \label{sec:appdx1}}
In this appendix, we provide the explicit expressions of the G-BSSN 
evolution equations in spherical symmetry. 
 
The evolution equations (\ref{eq:evolphi}-\ref{eq:evolctm}) for 
$\phi$ 
and the components of the conformal metric $\ctm$ simplify to

\def\dphi{\partial_r \phi}
\def\ddphi{\partial_r^2 \phi}
\def\dA{\partial_r \tilde{\gamma}_{rr}}
\def\ddA{\partial_r^2 \tilde{\gamma}_{rr}}
\def\dB{\partial_r \tilde{\gamma}_{\theta\theta}}
\def\ddB{\partial_r^2 \tilde{\gamma}_{\theta\theta}}
\def\dK{\partial_r K}
\def\dbeta{\partial_r \beta}
\def\ddalpha{\partial_r^2 \alpha}
\def\dalpha{\partial_r \alpha}
\def\pp{|\Pi|^2}
\def\hh{|\Phi|^2}
\def\pot{V(|\Psi|)}

\begin{equation}
\partial_t \phi = \frac{1}{6}\alpha K +\beta\dphi +\sigma \frac{1}{6}\Beta \, ,
\end{equation}
\begin{equation}
\partial_t \A = -2\alpha \tA_{rr} +\beta \dA +2\A\dbeta -\sigma\frac{2}{3}\A\Beta \, , 
\end{equation}
\begin{equation}
\partial_t \B = -2\alpha \tA_{\theta\theta} +\beta \dB +2\frac{\beta}{r}\B -\sigma\frac{2}{3}\B\Beta \, ,
\end{equation}
where $\Beta$ is the divergence of the shift vector,
\begin{equation}
\Beta(t,r) = D_{i}\beta^i = \dbeta + \frac{2\beta}{r} + \beta\left( \frac{\dA}{2\A} + \frac{\dB}{\B} \right) \, .
\end{equation}
To display the equation of motion for the trace of the extrinsic curvature $K$ and $\tilde{A}_{ij}$ we first define
\begin{equation}
\mathscr{D}_{ij} \equiv D_i D_j \alpha \, ,
\end{equation}
which has 2 independent components,
\begin{equation}
\mathscr{D}_{rr} = \ddalpha - \dalpha\left( \frac{\dA}{\A} + 4\dphi \right) \, ,
\end{equation}
\begin{equation}
\mathscr{D}_{\theta\theta} = r\dalpha \frac{\B}{\A} + \frac{r^2}{2}\dalpha\left(\frac{\dB}{\A} + 4\dphi\frac{\B}{\A}  \right) \, .
\end{equation}
The trace of $\mathscr{D}_{ij}$ is
\begin{equation}
\mathscr{D} \equiv \gamma^{ij}\mathscr{D}_{ij} = e^{-4\phi} \left( \frac{\mathscr{D}_{rr}}{\A} + 2\frac{\mathscr{D}_{\theta\theta}}{r^2\B} \right) \, .
\end{equation}
Then the evolution of $K$ is given by
\begin{eqnarray}
\partial_t K &=& -\mathscr{D} + \alpha\left( \frac{1}{3}K^2 + \frac{\tA_{rr}^2}{\A^2} + 2\frac{\tA_{\theta\theta}^2}{\B^2} \right) \nonumber \\
            &+&\beta \dK +4\pi\alpha(\rho+S) \, 
\end{eqnarray}
and the evolution equations for the traceless part of the extrinsic curvature are
\begin{eqnarray}
\partial_t \tA_{rr} &=& e^{-4\phi}\left[ -\mathscr{D}^{\mathrm TF}_{rr} +\alpha\left( R^{\mathrm TF}_{rr} + 8\pi S^{\mathrm TF}_{rr}  \right) \right] \nonumber \\
                  &+& \alpha\left( \tA_{rr}K - \frac{2\tA_{rr}^2}{\A}  \right) \nonumber \\
                  &+& 2\tA_{rr}\dbeta + \beta \partial_r \tA_{rr} -\sigma \frac{2}{3}\Beta \tA_{rr} \, ,
\end{eqnarray}
\begin{eqnarray}
\partial_t \tA_{\theta\theta} &=& \frac{e^{-4\phi}}{r^2} \left[  -\mathscr{D}^{\mathrm TF}_{\theta\theta} +\alpha\left( R^{\mathrm TF}_{\theta\theta} + 8\pi S^{\mathrm TF}_{\theta\theta}  \right) \right] \nonumber \\
                            &+& \alpha\left( \tA_{\theta\theta}K - 2\frac{\tA_{\theta\theta}^2}{\B}  \right) \nonumber \\
                            &+& 2\frac{\beta}{r}\tA_{\theta\theta} + \beta \partial_r \tA_{\theta\theta} - \sigma\frac{2}{3}\tA_{\theta\theta}\Beta \, ,
\end{eqnarray}
where $R$ denotes the 3-Ricci tensor with non-vanishing components
\begin{eqnarray}
R_{rr} &=& \frac{3(\dA)^2}{4\A^2} - \frac{(\dB)^2}{2\B^2} + \A\partial_r \tLambda + \frac{1}{2}\dA\tLambda \nonumber  \\ 
       &+& \frac{1}{r} \left(  4\dphi - \frac{\dA- 2\dB}{\B} - \frac{2\A\dB}{\B^2}  \right)  \nonumber \\
       &-&  4\ddphi + 2\dphi\left( \frac{\dA}{\A} - \frac{\dB}{\B} \right) \nonumber \\ 
		 &-& \frac{\ddA}{2\A}  + \frac{2(\A-\B)}{r^2\B} \, ,
	\label{eq:riccitenrr}
\end{eqnarray}
\begin{eqnarray}
R_{\theta\theta } &=& \frac{r^2\B}{\A}\left( \dphi\frac{\dA}{\A} -2\ddphi -4(\dphi)^2  \right) \nonumber \\
                  &+& \frac{r^2}{\A} \left( \frac{(\dB)^2}{2\B} - 3 \dphi\dB - \frac{1}{2} \ddB \right) \nonumber \\
                  &+& r \left( \Lambda \B - \frac{\dB}{\B} - \frac{6\dphi \B}{\A} \right) \nonumber \\ 
						&+& \frac{\B}{\A} - 1 \, .
\end{eqnarray}
In the above expressions
the superscript TF denotes application of the trace-free-part operator, whose action can be written explicitly as
\begin{equation}
	{X}^{{\mathrm TF}}_{rr} = {X}_{rr} - \frac{1}{3}\gamma_{rr}{X} = \frac{2}{3}\left( {X}_{rr} - \frac{\gamma_{rr}{X}_{\theta\theta}}{\gamma_{\theta\theta}r^2} \right) \, ,
\end{equation}
\begin{equation}
	{X}^{{\mathrm TF}}_{\theta\theta} = {X}_{\theta\theta} - \frac{1}{3}\gamma_{\theta\theta}{X} = \frac{1}{3}\left( {X}_{\theta\theta} - \frac{\gamma_{\theta\theta}{X}_{rr}}{\gamma_{rr}} \right) \, .
\end{equation}
Here $X$ represents any of the tensors $\mathscr{D}$, $R$ or $S$.

Finally, the evolution of $\tilde{\Lambda}^i$ reduces to
\begin{eqnarray}
  \partial_t \tLambda &=& \beta\partial_r\tLambda - \dbeta\tLambda  +\frac{2\alpha}{\A} \left( \frac{6\tA_{\theta\theta} \dphi}{\A} - 8\pi S_r - \frac{2}{3}\dK  \right) \nonumber \\
  &+& \frac{\alpha}{\A}\left( \frac{\dA \tA_{rr}}{\A^2} - \frac{2\dB \tA_{\theta\theta}}{\B^2} + 4\tA_{\theta\theta}\frac{\A-\B}{r\B^2}  \right) \nonumber \\
  &+&  \sigma\left(\frac{2}{3}\tLambda\Beta + \frac{\partial_r\Beta}{3\A}\right) + \frac{2}{r\B} \left( \dbeta - \frac{\beta}{r} \right) \nonumber \\
  &-& 2\frac{\dalpha \tA_{rr}}{\A^2} \, .
\end{eqnarray}
\section{Scalar field dynamics and energy-momentum tensor in spherical symmetry\label{sec:appdx2}}
Here we present the evolution equations of a complex scalar field, with
an arbitrary potential $V$, minimally
coupled to gravity. The governing equations for a massless real scalar field follow as a 
special case
where the imaginary part of the field and the potential are both set to zero.

The geometry of spacetime is given by a generic metric 
in spherical symmetry:
\begin{equation}
	ds^2 = (-\alpha^2+\beta^2a^2)dt^2 + 2a^2\beta dtdr + a^2dr^2 + r^2b^2d\Omega^2 ,
\end{equation}
where $a$, $b$, $\alpha$ and $\beta$ are all functions of $t$ and $r$ and 
where $a$ and $b$ are related to the primary BSSN variables via
$a=\A\exp(4\phi)$ and $b=\B\exp(4\phi)$. 

The complex scalar field is given in terms of 
real and imaginary parts,
$\Psi_R$ and $\Psi_I$, respectively,
\begin{equation}
\Psi = \Psi_{R}(t,r) + i\Psi_{I}(t,r) \, ,
\end{equation}
and has an associated energy-momentum tensor
\begin{eqnarray}
T_{\mu\nu} &=& \nabla_{\mu}\nabla_{\nu}\Psi_R -\frac{1}{2}g_{\mu\nu}\nabla^{\eta}\Psi_R\nabla_{\eta}\Psi_R \nonumber \\
           &+&\nabla_{\mu}\nabla_{\nu}\Psi_I -\frac{1}{2}g_{\mu\nu}\nabla^{\eta}\Psi_I\nabla_{\eta}\Psi_I \nonumber \\
			&-& \frac{1}{2}g_{\mu\nu}\pot \, .
\end{eqnarray}
The evolution of the real part of the scalar field can be reduced to a pair of first-order-in-time
equations via the definition
\be
\Xi_R \equiv \frac{b^2 a}{\alpha}\left( \partial_t \Psi_R - \beta \partial_r \Psi_R \right) \, .
\ee
We then find the following evolution equations for $\Psi_R$ and $\Xi_R$:
\be
\partial_t \Psi_R = \frac{\alpha}{b^2a} \Xi_R + \beta \partial_r \Psi_R \, ,
\label{eq:dtPsiR}
\ee
\begin{eqnarray}
\partial_t \Xi_R &=& \frac{\alpha b^2}{a} \left( \partial_r^2 \Psi_R + 2\frac{\partial_r\Psi_R}{r} \right) + \partial_r\Psi_R \partial_r\left( \frac{\alpha b^2}{a}  \right) \nonumber \\
                 &+& \beta\partial_r\Xi_R + \Xi_R\partial_r\beta + \Xi_R\frac{2\beta}{r} \nonumber \\
                 &-& a\alpha b^2 \partial_{|\Psi|}^2\pot \, .
\label{eq:dtXiR}
\end{eqnarray}
The evolution equations for $\Psi_I$ and $\Xi_I$ follow from the index substitutions 
$R\leftrightarrow I$ in the right hand sides of (\ref{eq:dtPsiR}) and~(\ref{eq:dtXiR}), 
respectively.

The matter source terms in the G-BSSN equations, namely $\rho$, $S$, $S^i$, $S_{ij}$, can
be simplified by defining $\Pi$ and $\Phi$ as
\be
\Pi  \equiv \frac{a}{\alpha} \left( \partial_t \Psi - \beta \partial_r \Psi \right) \equiv \Pi_R(t,r) + i\Pi_I(t,r) \, ,
\ee
\be
\Pi_{R} = \frac{a}{\alpha} \left( \partial_t\Psi_R -\beta\partial_r\Psi_{R} \right) =  \frac{\Xi_R}{b^2} \, ,
\ee
\be
\Pi_{I} = \frac{a}{\alpha} \left( \partial_t\Psi_I -\beta\partial_r\Psi_{I} \right) = \frac{\Xi_I}{b^2} \, ,
\ee
\be
\Phi  \equiv \partial_r\Psi \equiv \Phi_R(t,r) + i\Phi_I(t,r) \, ,
\ee
\be
\Phi_{R} = \partial_r \Psi_R \, ,
\ee
\be
\Phi_{I} = \partial_r \Psi_I \, .
\ee
Using these definitions, the variables $\rho$ and $S$ are given by
\be
\rho(t,r) = \frac{\pp+\hh}{2a^2} + \frac{\pot}{2} \, ,
\ee
\be
S(t,r) = \frac{3\pp -\hh}{2a^2} - \frac{3}{2}\pot \, .
\ee
In spherical symmetry, $S^i$ has only a radial component,
\be
S^i = \left[S^r(t,r),0,0\right] \, ,
\ee
with
\be
S^r = -\frac{\Pi_R\Phi_R + \Pi_I\Phi_I}{a} \, .
\ee
Similarly, the spatial stress tensor, $S_{ij}$, has only two independent components,
\be
S_{ij} = \dtm{S_{rr}(t,r)}{r^2S_{\theta\theta}}{r^2\sin^2\theta S_{\theta\theta}} \, ,
\ee
with
\be
S_{rr} = \frac{\pp +\hh}{2} - a^2\frac{\pot}{2}\, ,
\ee
\be
S_{\theta\theta} = b^2 \left( \frac{\pp -\hh}{2a^2}  - \frac{\pot}{2}  \right) \, .
\ee


\bibliography{paper}

\end{document}